\documentstyle[prb,preprint,aps] {revtex}
\begin{document}

\draft
\preprint{Submitted to PRB}
\title{
Nonbackscattering Contribution to Weak Localization
}
\author{A.P. Dmitriev \cite{ADVK} and
V.Yu.Kachorovskii \cite{ADVK}}
\address{Uppsala University, S-751 08, Uppsala, Sweden}
\author{I.V. Gornyi}
\address{
A.F.Ioffe Physical-Technical Institute, St.Petersburg, 194021,
Russia}
\maketitle

\begin{abstract}

    We show that the enhancement of backscattering
responsible for the weak localization is accompanied by
reduction of the scattering in other directions. A simple
quasiclassical interpretation of this phenomenon is
presented in terms of a small change in the effective
differential cross-section for a single impurity. The
reduction of the scattering at the arbitrary angles leads
to the decrease of the quantum correction to the
conductivity. Within the diffusion approximation this
decrease is small, but it should be taken into account in
the case of a relatively strong magnetic field when the
diffusion approximation is no more valid.
\end{abstract}
\pacs{PACS numbers: 73.20.Fz, 72.20.Dp, 72.10.-d }

\def \br {{\bf r}}

\newpage

\section { Introduction}
\label{sec:Intr}

The quantum correction to the conductivity arises from interference
of electron waves propagating in opposite
directions along closed paths. The interference is
destroyed for trajectories which are long enough.
In the absence of magnetic field and if spin effects may be
neglected it
happens due to processes of electron inelastic scattering
which are usually
taken into account by introducing  the
phase breaking time $\tau_{\phi}$. At sufficiently low
temperatures $\tau_{\phi}$ is much greater than the
elastic scattering time $\tau$
and the motion of electrons may be described by
a diffusion equation (diffusion approximation).

The corresponding conductivity correction is negative and
in the two
dimensional case is given by \cite{glk}:

\begin{equation}
\Delta \sigma=-\frac{e^2}{2\pi^2\hbar}\ln{\frac{L_\phi^2}{l^2}}.
\label{f1}
\end{equation}

Here  $L_\phi=({2D\tau_\phi})^{1/2}$ is the phase breaking length,
$D=l^2/(2\tau)$ is the diffusion coefficient and $l$ is the
mean free path.

It is well known \cite{aakl}, that Eq. (\ref{f1}) allows
a simple quasiclassical derivation based on calculation of the
probability for an electron
to return to the origin.

The presence of magnetic field leads to the phase coherence
distortion when the path linear dimensions are larger
than the magnetic length $l_H=(\hbar c/{eB})^{1/2}$. With increasing
magnetic field, $B$,  the magnetic length becomes
smaller than  $L_\phi$ and, accordingly, the conductivity
correction decreases
\cite{hln}.
For relatively weak magnetic fields, when $l\ll l_H \ll L_\phi$,
the equation (\ref{f1}) is still valid with $L_\phi$ being
changed by the length of the order of $l_H$.
For stronger magnetic fields when  $l_H \ll l$ (but still $l \ll
R_c$, $R_c$ is the cyclotron radius ),
the main contribution to the conductivity correction
comes from  short  closed trajectories with the length of the order of
$l_H$ and the diffusion approximation is no more valid.
This case was considered in the Refs. \onlinecite{zuzi,dyak}
and it was found that in two dimensions
for short range potential
$\Delta \sigma\propto -l_H/l$.

The quantitative theory of weak localization is based on the
expansion of the conductivity in series of the small parameter
$(k_F l)^{-1}$, where $k_F$ is Fermi wave vector.
The negative correction to the conductivity Eq. (\ref{f1}) 
arises in the first order of this parameter. 
It can be derived by summing so-called
maximally-crossed diagrams (Fig. 1a). These diagrams
describe the  coherent backscattering of the electron wave.
In the case when the diffusion approximation is not valid,
together with the diagrams (1a)
one should also take into account the diagrams presented in
Figs. 1b, 1c and 1d.  These diagrams too, give a contribution to the
conductivity of the order of  $(k_F l)^{-1}$
but, in contrast to the diagrams (1a), their contribution is
positive.
The importance of these diagrams were
emphasized in many works, but a clear  quasiclassical interpretation
of processes corresponding to these diagrams was never given.
Moreover in Ref. \onlinecite{baran} it was claimed that
a quasiclassical interpretation of these processes is not
possible.
In this work we present
a simple quasiclassical interpretation of any diagram
of the first order in  $(k_F l)^{-1}$.
It is shown that the contribution of these diagrams may be
expressed through the classical probability for an electron
to return to the
origin at a certain angle to the initial direction of motion.

    We discuss the possibility of describing weak
localization effects
in terms of a small change of the differential cross-section
of a single impurity.  The angular dependence of this modified
cross-section  for the case of zero
magnetic field and the short-range impurity potential
is presented in Fig. 2.
The positive peak near
$\theta=\pi$ corresponds to the enhancement of
backscattering described by the diagrams (1a)
while the other diagrams in Fig. 1 are responsible for
the decrease of the scattering rate in other directions, the total
cross-section remaining unchanged.
At the same time the transport cross-section changes and this is
the reason for the weak localization corrections.
This means that all first order in
$(k_F l)^{-1}$ weak localization effects
may be taken into account by changing the
differential cross-section of a single impurity.
A similar consideration is also possible when magnetic
field is applied. In this case the effective cross-section
depends on magnetic field.

  It is also shown that within the diffusion approximation ($L_\phi,
l_H\gg l$) taking into account
the diagrams (1b,c and d)
leads to the appearance in Eq. (\ref{f1}) of an additional
factor $1/2$ in the argument of the logarithm. At strong
magnetic fields ($l_H<l$),
when the diffusion approximation is no longer valid,
the contribution of the diagrams (1b,c and d) differs from
that of the diagrams (1a) by the numerical factor only.

We calculate numerically the quantum correction to the
conductivity for the total range of the classically weak magnetic
fields. The results are presented graphically.

The paper is organized as follows.
In the first section we give the necessary formulas and
definitions.
In the second section the derivation of the correction to the
conductivity due to the diagrams (1a) 
is given in the coordinate representation. This
method allows to reach more transparent physical presentation.
In the third section the quasiclassical interpretation of the
rest diagrams in Fig. 1 is given, using the same method. The
dependence of the quantum correction on the magnetic field is
considered.
Finally, in the fourth section we discuss the possibility
of describing the weak localization in terms of
interference correction to the
differential cross-section.

\section {Basic equations}
\label{sec:ME}

We consider the motion of 2D-electrons in a random potential
 $V(\br)=\sum u(\br-\bf R_i),$
where $\bf R_i$ is a vector of the position of
$i$'th impurity, $u(\br)$ is a single impurity potential, which is
supposed to be a short-range one.
The correlation function of the total potential $V(\br)$ is
given by:
\begin{equation}
\langle V(\br) V(\br^\prime) \rangle = \gamma \delta(\br-\br^\prime).
\label{f2}
\end{equation}
Here the angular brackets denote averaging over the
impurity's positions.
Static conductivity is calculated with the use of the Kubo
formula. It will be convenient for our purposes to write this formula
in the coordinate representation:
\begin{equation}
\sigma = - \frac{e^2 \hbar^3}{2\pi m^2 S}
\int\int d^2 \br_i d^2 \br_f\left \langle \frac{\partial}{\partial \br_i}
 G^R_e (\br_i,\br_f,E_F)
 \frac{\partial}{\partial \br_f} G^A_e (\br_f,\br_i,E_F) \right \rangle.
\label{f3}
\end{equation}
Here $m$ is the electron mass, $S$ is the area of the system,
$G^{R,A}_e (\br,\br^\prime,E_F)$
are  respectively the retarded and advanced exact Green functions
at the Fermi energy $E_F$.

As is well known, the result of averaging over impurity's
positions is represented
as a sum of all the possible diagrams with
solid lines corresponding to averaged Green functions and dashed
lines corresponding to the potential correlation function.

The expressions for the averaged Green functions are given by
\begin{equation}
G^{R,A} (\br,\br^\prime,E_F)=
\left \langle  G^{R,A}_e (\br,\br^\prime,E_F)\right \rangle=
\int \frac {d^2 {\bf k} \exp (i{\bf k}(\br-\br^\prime))}
{\displaystyle (2\pi)^2(E_F-\frac{\hbar^2 k^2}{2m} \pm \frac{i\hbar}{2\tau})},
\label{f4}
\end{equation}
where $ \tau={\hbar^3}/{(m \gamma)}$ is the elastic 
scattering time \cite{snoska}.
These expressions have the following asymptotic behavior at
distances exceeding the wave length
\begin{equation}
G^{R,A} (\br-\br^\prime,E_F)=
\mp \frac{ i m}{\hbar^2}
\frac {1}{\sqrt{2\pi k_F |\br-\br^\prime|}}
\exp({\pm ik_F |\br-\br^\prime|-
\frac{|\br-\br^\prime|}{2 l} \mp i\frac{\pi}{4}}).
\label{f5}
\end{equation}
The Green functions $G^R$ and $G^A$ describe the divergent and
convergent waves respectively. These waves oscillate
rapidly on the scale $k_F^{-1}$ and their amplitudes decrease slowly on the
scale of the order of the mean free path $l$.
The large value of the parameter
$k_F l$ allows to give a quasiclassical interpretation for various
terms in the diagram series, the quantity
\begin{equation}
P(\br)=\gamma G^R(\br,E_F) G^A(\br,E_F) =
\frac {e^{-r/l}}{2\pi l r}
\label{f6}
\end{equation}
playing an essential role. This is a classical probability density
for an electron starting from the origin $\br=0$ to
experience the first collision around point  $\br$.

In what follows we will make use of the relation
\begin{equation}
\int d^2\br_iG^A_{iN}\frac{\partial}{\partial \br_i}G^R_{i1}=
\frac{i\tau}{\hbar}\frac{\partial}{\partial \br_N}
(G^R_{N1}-G^A_{N1})\approx
-\frac{ml}{\hbar^2} \frac{(\br_N-\br_1)}{|\br_N-\br_1|}
\left[G^R_{N1}+G^A_{N1}\right],
\label{f9}
\end{equation}
which may be easily derived from  Eq. (\ref{f4}).
Here we use the notation  $G^{R,A}_{jk}=G^{R,A}(\br_j-\br_k)$.

For a short-range potential, when the scattering is isotropic, the
main contribution to the conductivity is given by the diagram
without dashed lines, which corresponds to independent
averaging of the Green functions in
 (\ref{f3}). It is easy to see
that in this approximation Eq. (\ref{f3}) is reduced to the
integral
\begin{equation}
\sigma_0=\frac{e^2 \tau k_F^2}{2\pi m}\int P(\br_i-\br_f)
\frac{d^2\br_i d^2\br_f}{S}
=\frac{e^2 n \tau}{m}=\frac{e^2}{2\pi \hbar}k_F l,
\label{f7}
\end{equation}
where $n$ is the electron concentration.
This equation is in fact the classical Drude formula

\section {  Coherent backscattering correction}
\label{sec:CBC}

The coherent backscattering correction to the Drude formula
(\ref{f7}) is described, in the first order in $(k_Fl)^{-1}$,
by the diagrams (1a),
the number of dashed lines being greater than two \cite{two}.

These diagrams represent the contribution to conductivity
related to interference of two processes depicted in Fig. 3a.
An electron starting from the point $i$
reaches the point $f$ by two ways:

1) successively scattering on impurities $1,2,...,N$ ,

2) passing the same chain of impurities in the opposite order
($N,N-1,...1$). \\
It means that each section of the trajectory from $1$ to $N$
is passed twice.
The amplitudes of these transitions are described by the
functions
$G^R$ and $ G^A$ respectively
which come into the expression for  the conductivity correction
as  products $\gamma G^R(\br_j-\br_{j+1})G^A(\br_j-\br_{j+1})=
P(\br_j-\br_{j+1})$.
Thus the phase difference of the two waves on the paths connecting
points $1$ and $N$ is equal to zero and
the quantity
\begin{equation}
W_{N-1}(\br_1-\br_N)=\int d^2\br_2...d^2\br_{N-1}
P(\br_1-\br_2)...P(\br_{N-1}-\br_N)
\label{f8}
\end{equation}
appears in the expression for the conductivity correction.
This quantity is the
classical probability density to find an
electron started from point $1$ near the point $N$
after  $N-1$ collisions.

The smallness of the contribution to conductivity of the
diagrams (1a) in comparison with the main Drude's one (Eq. (\ref{f7}))
results from the initial and last sections of the trajectories
$(i,1)$,  $(i,N)$ and $(N,f)$, $(1,f)$
that normally are passed only once (see Fig. 3a).
The total phase difference of the two waves at
point $f$ comes from these sections only and
is given by
\begin{equation}
\Delta\phi=k_F(|\br_1-\br_i|+ |\br_f-\br_N|-|\br_N-\br_i|-|\br_f-\br_1|).
\label{faza}
\end{equation}
The smallness arises after
integrating over the coordinates of the points
$i$ and $f$ in Eq. (\ref{f3}), due to rapid
oscillations of $\exp (i\Delta\phi)$.
The main contribution to the integral comes from
such configurations for which the phase difference is
stationary with respect to small variations of the coordinates
of all four key points ($i$, $1$, $N$ and $f$).
This happens when all these points are close to
one line, the points $i$ and $f$ lying
on the one side from the section $1$--$N$ (see Fig. 3b).
That is why the processes described by the diagrams (1a)
may be interpreted as an additional backscattering on a
single impurity (the impurity $1$ for the configuration
depicted in Fig. 3b).

We stress that it is the condition that the
phase difference $\Delta\phi$ be stationary that
is important, but not the condition $\Delta\phi=0$.
There are configurations for which $\Delta\phi=0$,
but stationarity condition is not valid (for example,
when the points $i$ and $f$ lie symmetrically
with respect to the line $1-N$). Such configurations
do not contribute to the quantum correction. It turns out
however, that in the case presented in Fig. 3b the
total phase difference is
equal to zero and constructive interference takes place.

The coherent backscattering correction to conductivity
can be expressed through the classical probability
density for an  electron to return to the area of the order
$l\lambda_F$ ($\lambda_F=2\pi/k_F$) around
the impurity $1$ (see Appendix A):
\begin{equation}
\Delta \sigma_a=-\sigma_0 \frac{(\lambda_F l)}{\pi} W.
\label{coop}
\end{equation}
Here
\begin{equation}
W=\sum^{\infty}_{N=3} W_N(0)
\label{f10}
\end{equation}
is the sum of probability densities for an electron to return
to the origin after $3,4,..$ collisions. In what follows,
for the sake of brevity
we will name this quantity as the total probability of return
\cite{chak}.

It is easy to see that
\begin{equation}
W=\int\frac{d^2k}{(2\pi)^2}\frac{P_k^3}{1-P_k}.
\label{fff}
\end{equation}
Here the quantity
 $P_k=(k^2l^2+1)^{-1/2}$
is the Fourier-transform of $P(\br)$.

The fact that electron should return to the area $\lambda_F l$
around the impurity $1$ can be explained in the following way.
The distance between points $1$ and $N$ should be of the order of
$l$ in consequence of waves fading on the mean free path.
Thus only paths which pass at a distance
$(\lambda_Fl)^{1/2}$ from impurity $1$ (see Fig. 3b)
interfere.

Without taking into account the inelastic processes
the integral in (\ref{f10}) diverges logarithmically.
In order to take into account such processes
one can replace $1/\tau$ by $(1/\tau+1/\tau_\phi)$
in the Eq. (\ref{f4}).
Then the quantity $P_k$ is given by
\begin{equation}
P_k=\frac{1}{\sqrt{k^2l^2+(1+{\tau}/{\tau_\phi})^2}}.
\label{f11}
\end{equation}

After integrating in (\ref{f10}) we finally obtain
\begin{equation}
\Delta \sigma_{a}=
-\frac{e^2}{2\pi^2\hbar}\ln{\frac{\tau+\tau_{\phi}}{\tau}}.
\label{f12}
\end{equation}

This formula represents the coherent backscattering correction
to conductivity.

\section { Correction to the conductivity due to scattering at
arbitrary angle }
\label{sec:nasha}

The set of diagrams which describe the
corrections to conductivity of the order of
$(k_Fl)^{-1}$
is not restricted by the series of diagrams (1a) only.
The diagrams presented in Figs. 1b, 1c and 1d should also be taken
into account.
In the absence of magnetic field the contributions
of such diagrams
to the conductivity are of the same absolute value
but differ in sign.
The contribution of the diagrams of Figs. 1b and 1c is positive
whereas the contribution of the diagrams in Fig. 1d is negative.
It is easy to show, that magnetic field does not
change the contributions of diagrams in Fig. 1c and Fig. 1d
and they still compensate each other.  Thus, when calculating the
correction to conductivity one should take into account only the
diagrams in Fig. 1b,
both in the presence and in the absence of magnetic field .

Let us show that the process described by diagrams (1b)
can be easily interpreted quasiclassically (the diagrams in
Figs. 1c and 1d allow a similar interpretation). Such a process
is depicted in Fig. 4a.
An electron starting from point $i$
reaches point $f$ by two ways:

1) consecutively scattering by impurities $1,2,...,N$
and finally by impurity $1$ again,

2) scattering in the opposite order by impurities
$N,N-1,..,2$, and
having no collisions at all with impurity $1$.\\
The classical quantities $P(\br_j-\br_{j+1})$ not
containing phase factors correspond to the intervals
$(2,3),(3,4),..,(N-1,N)$.
The integration over the coordinates of impurities
$3,...,N-1$ leads to the appearance of the function
$W_{N-2}(\br_N-\br_2)$.

The phase difference of the two paths
ending at the point $f$ depends on the lengths of the intervals
 $(i,1),(1,2),(N,1),(1,f)$ and $ (i,N),(2,f)$.
and is given by

\begin{equation}
\Delta\phi=k_F(|\br_1-\br_i|+|\br_2-\br_1|+ |\br_1-\br_N|
+|\br_f-\br_1|-|\br_N-\br_i|-|\br_f-\br_2|).
\label{geom}
\end{equation}

Let us fix the positions of the points $i,1,f$
and then integrate over the coordinates of
the impurities $2, N$.
Because of the  phase stationarity requirement
the contribution to the conductivity arises only from
the configurations in which the points $N$ and $2$
lie close to the lines $i$--$1$ and $1$--$f$ respectively
in angles of the order of $(k_F l)^{-1/2}$ (see Fig. 4b).
In this configuration $\Delta\phi$ is equal to zero.
It is clear from Fig. 4b that the process described by
diagrams (1b) can be interpreted as a coherent changing
of the scattering by the impurity $1$ at angle
$\theta$. It can be shown that a reduction of scattering
takes place. \cite{pi}

The expression for the conductivity correction due to processes
of Fig. 4b can be written as (see Appendix B):

\begin{equation}
\Delta \sigma_{b}= -\sum^{\infty}_{N=3}
\frac{\sigma_0}{S\pi} (\lambda_F l)
\int d^2\br_1d^2\br_2d^2\br_N P(\br_1-\br_2)W_{N-2}(\br_N-\br_2)
P(\br_N-\br_1)\cos\theta ,
\label{nasha}
\end{equation}
$$
\cos\theta=\frac{(\br_N-\br_1)(\br_1-\br_2)}{|\br_N-\br_1||\br_1-\br_2|}.
$$

Using the Fourier transformation one can get
\begin{equation}
\Delta \sigma_{b} = \frac{e^2}{\pi\hbar}l^2
\int\frac{d^2k}{(2\pi)^2}\frac{P_k(P_k^\prime)^2}{1-P_k},
\label{p1}
\end{equation}
where $P_k^\prime=(1/kl)(1-P_k)$ is the Fourier component
of the function
$-i P(\br)\cos\alpha$, $\alpha$ is the angle of the vector $\br$.
Calculating the integral in Eq. (\ref{p1}), we finally obtain:
\begin{equation}
\Delta \sigma_{b}=\frac{e^2}{2\pi^2\hbar}\left(
\frac{\ln2}{1+\tau/2\tau_\phi}-\frac{\ln(1+\tau_\phi/\tau)}
{1+2\tau_\phi/\tau}\right).
\label{nasha1}
\end{equation}
Note that this correction is
positive in contrast to contribution due to the
coherent backscattering.
In the diffusion approximation ($\tau_\phi\gg\tau$)
the expression (\ref{nasha1}) simplifies:
\begin{equation}
\Delta \sigma_{b}=\frac{e^2}{2\pi^2\hbar}\ln 2.
\label{ln2}
\end{equation}
The total (with accounting both (1a) and (1b) diagrams)
weak localization correction to conductivity
in the diffusion approximation is given by
\begin{equation}
\Delta \sigma=-\frac{e^2}{2\pi^2\hbar}\ln(\frac{{L_{\phi}}^2}{2l^2}).
\label{total}
\end{equation}

Thus when the diffusion approximation is valid
the contribution of the diagrams (1b) is logarithmically
small compared
to the backscattering one and just leads to the appearance of
a factor $1/2$ in the argument of the large logarithm.

Beyond the diffusion approximation, when only the
trajectories with a small number of collisions are
important, the situation is quite
different. This happens in sufficiently strong magnetic
field when the magnetic length $l_H$ is of the order of the
mean free path $l$, or less.
In this case the correction arising from diagrams (1a) 
does not contain
the large logarithm and contributions of the diagrams (1b)
and the diagrams (1a) differ only by a numerical factor
of the order of unity.

In the presence of magnetic field,
Eqs. (\ref{f8}),
(\ref{coop}),(\ref{f10}) and (\ref{nasha}) still
hold, but the quantity
$P(\br-\br^\prime)$ should be replaced by
 $$
\tilde P(\br-\br^\prime)=P(\br-\br^\prime)
\exp(i(e/\hbar c){\bf B}[\br \times \br^\prime]).
$$

Using Kawabata's method \cite{kawa} one can expand the
functions   $\tilde P(\br),\:W(\br)$ in terms of the eigenfunctions
of a particle of charge  $2e$ in a magnetic field  $B$
and obtain

$$
\Delta \sigma
= - \frac {e^2}{2\pi^2 \hbar}F(x),\quad
F(x)=F_{a}(x)+F_{b}(x),
$$
$$
F_{a}(x)=x\sum^{\infty}_0\frac{(P_n)^3}{1-P_n},\quad
F_{b}(x)=-x\sum^{\infty}_0\frac{P_n((P_n^1)^2/2+(P_n^{-1})^2/2)}{1-P_n},
$$
where
$$
P_n=\frac{s}{z}\int^{\infty}_0dtexp(-st-t^2/2)L_n(t^2),
$$
$$
P_n^m=\frac{s}{z\sqrt{n+(1-m)/2}}
\int^{\infty}_0dtexp(-st-t^2/2)L_n^m(t^2),
$$
$x=B/B_0$, $B_0=\hbar c/(2el^2)$, $s= z(2/x)^{1/2}$,
$z=1+\tau/\tau_{\phi}$,
$L_n$ and $L_n^m$ are the Laguerre polynomials.
The functions $F_{a}(x)$ and $F_{b}(x)$ describe the
contributions of diagrams (1a) and (1b) respectively.
In the high-field limit the quantum correction to
conductivity has the form\cite{zuzi,dyak}
$$\Delta \sigma = \Delta\sigma_a+\Delta\sigma_b 
= -4.96 \frac{e^2}{2\pi^2\hbar}\frac{1}{\sqrt{x}},
$$
$$
\Delta\sigma_a=-7.74\frac{e^2}{2\pi^2\hbar}\frac{1}{\sqrt{x}},\quad
\Delta\sigma_b=2.78\frac{e^2}{2\pi^2\hbar}\frac{1}{\sqrt{x}}.
$$
Note, that this asymptotical behavior is valid only at
very high values of $x$ and can be hardly observed
in experiment.

We have performed numerical calculations of $\Delta \sigma(B)$
for all range of magnetic fields. The dependencies of
$\Delta\sigma(B)$ and $\Delta \sigma_{a}(B)$
for $\tau_\phi=\infty$ are presented in  Fig. 5.
The dependence $\Delta\sigma(B)$
for different values of $\tau_\phi$
is represented in Fig. 6.

\section {  Interpretation of the weak localization in terms of
changing of impurity scattering cross-section
}
\label{sec:interpr}

The method presented above allows to give a transparent
interpretation of weak localization effects.

In Refs. \onlinecite{kirk,ambe} the
effects, described by the diagrams (1a) were treated 
in the frame of the Boltzman transport equation.
The authors of Ref. \onlinecite{ambe} claim that the main weak
localization effect is an effective reduction of elastic
scattering time. 

Using the ideas of these works it is easy to
show that the processes related to diagrams (1b) can be
considered in the frame of the Boltzman transport equation
as well as coherent backscattering processes.
One should just replace the isotropic cross-section
$S_0$ by the following expression:
\begin{equation}
S(\theta)=S_0+\Delta S(\theta).
\label{cross}
\end{equation}
Here the function $S(\theta)$ is the modified impurity
scattering cross-section
which is represented schematically in Fig. 2 and
$$
\Delta S(\theta)=\Delta S_a(\theta)+\Delta S_b(\theta).
$$
The term  
$$
\Delta S_a(\theta)=C\Delta(\pi-\theta)
$$
corresponds to the coherent backscattering
at small angles of the order of $(k_F l)^{-1}$. The function
$\Delta(\pi-\theta)$ is concentrated in this angle and
the integral of it over  $\theta$ is equal to unity.
The quantity $C$ is expressed through the total probability of
return
$W$:
\begin{equation}
C=\frac{S_0}{k_F l} 4\pi l^2 W.
\label{ccc}
\end{equation}
In the diffusion approximation
 $W=\ln(\tau_\phi/\tau)/(2\pi l^2)$.

The function $\Delta S_b(\theta)$ is negative and
corresponds to a decrease of scattering at angle
$\theta$, being described by the diagrams in Fig. 1b.
 This function can be expressed through the total probability $W(\theta)$
for an electron to return to the origin at an angle $\theta$
to the initial direction of propagation

\begin{equation}
\Delta S_b(\theta)=-\frac{S_0}{k_F l} 4\pi l^2 W(\theta).
\label{s1}
\label{nashesech}
\end{equation}
The return probability $W(\theta)$ is given by
\begin{equation}
W(\theta)=2\pi \int rdr r^\prime dr^\prime P(\br) P(\br^\prime)
\sum_{N=1}^{\infty}W_{N}(|\br-\br^\prime|).
\label{dublwtheta}
\end{equation}
The integration in this equation should be done over
absolute values of vectors
$\br, \br^\prime $, the angle between them being fixed and equal to
$\pi-\theta $. For $\tau_{\phi}\gg \tau$ the
straightforward calculation gives
\begin{equation}
W(\theta)=\frac{1}{(2\pi)^2 l^2}\left[\ln\frac{\tau_\phi}{\tau}-
\ln\left|\cos{\frac{\theta}{2}}\right|-\frac{\pi-|\pi-\theta|}{2}
\cot{\frac{\theta}{2}}\right].
\label{trit}
\end{equation}
This expression is correct for
$|\pi-\theta|>(k_Fl)^{-1}$. In the opposite case
$\cos(\theta/2)$ in the second term
should be replaced by the quantity of the order of
$(k_Fl)^{-1}$. Within the diffusion approximation the main contribution to
$W(\theta)$ comes from the first term in Eq. (\ref{trit})
and therefore this function is almost isotropic. 
The anisotropic part of $W(\theta)$ arises mainly due to
triangle trajectories.

    It is easy to see from Eqs.
(\ref{ccc}),(\ref{nashesech}) and (\ref{dublwtheta})
that
\begin{equation}
\int^{2\pi}_0 W(\theta)d\theta=W,\qquad
\int^{2\pi}_0 \Delta S(\theta)d\theta=0.
\label{nol}
\end{equation}
This means, in contrast to the statement in Ref. \onlinecite{ambe},
that the weak localization effects do not change  the elastic
scattering time, which is inversely proportional to the
total cross-section.
The reduction of this time due to the coherent backscattering is exactly
compensated by its enhancement due to the reduction of the
scattering at other angles. This happens due to the fact
that each impurity configuration, contributing to coherent backscattering,
gives the contribution of the same value to scattering  
in angle $\theta$ too (see Figs. 3b and 4b).
 At the same time,
since the differential cross-section is
anisotropic due to the quantum corrections (see Fig. 2), the transport
scattering time
changes and does not anymore equal to the elastic
scattering time.
This is the physical reason which leads to the quantum
corrections to conductivity of the order of
$(k_Fl)^{-1}$.

We want to emphasize that the correct treatment of weak
localization effects in the framework of the Boltzman
equation 
is only possible when the diagrams (1b) are taken into
account. It can be explained by the following way.
For inverse transport scattering time we have
\begin{equation}
\frac{1}{\tau_{tr}}=\frac{1}{\tau S_0}
\int^{2\pi}_0 S(\theta)(1-\cos{\theta})d\theta=
\frac{1}{\tau}+\nu,
\label{tautr} 
\end{equation}
where the correction $\nu$ arises due to the term 
$\Delta S(\theta)$ in Eq. (\ref{cross}).
In the first order in $(k_Fl)^{-1}$ the transport
scattering time reads $\tau_{tr}=\tau(1-\tau\nu)$. Then 
for the conductivity we get the following
expression 
\begin{equation}
\sigma=\sigma_0
\left[1-\frac{1}{S_0}\int^{2\pi}_0 
\Delta S(\theta)(1-\cos{\theta})d\theta\right].
\label{krug} 
\end{equation} 

It is easy to see that taking into account in this method only the
contribution of diagrams (1a) (i.e., assuming that $\Delta
S(\theta)=\Delta S_a(\theta)$ ) 
leads to the conductivity correction 
which is twice greater than the correct one.

Using Eq. (\ref{nol}) we obtain for the conductivity correction
\begin{equation}
\Delta\sigma=-\frac{\sigma_0}{2\pi S_0}
\int^{2\pi}_0\Delta S(\theta)\cos{\theta}d\theta
=\Delta\sigma_a+\Delta\sigma_b,
\label{fin} 
\end{equation}
where 
\begin{equation}
\Delta\sigma_{a,b}=\frac{\sigma_0}{2\pi S_0}
\int^{2\pi}_0\Delta S_{a,b}(\theta)\cos{\theta}d\theta.
\label{finab} 
\end{equation}
These expressions for $\Delta\sigma_{a,b}$ coincide
with that derived by using the Kubo formula \cite{valya}.
Note, that after integrating in Eq. (\ref{finab}) the
isotropic part of $\Delta S_b(\theta)$ arising from the
first term in Eq. (\ref{trit}) does not contribute
to the conductivity. As a result the conductivity 
correction due to diagrams (1b) does not contain 
a divergent with $\tau_{\phi}$ contribution.  

    In the presence of magnetic field Eq. (\ref{cross})
remains valid. The quantities $W$ and $W(\theta)$
entering Eqs. (\ref{ccc}) and (\ref{s1})
should be calculated in this case using Eqs. (\ref{f8}), (\ref{f10}) and
(\ref{dublwtheta}) in which $ P(\br)$ should be replaced
by $\tilde P(\br)$.  In the high-field limit only triangle
paths are important and $W(\theta)$ is strongly
anisotropic and conductivity corrections
$\Delta\sigma_{a,b}$ differ by numerical factor of the
order of unity.

\section{Acknowledgments}
\label{sec:Acknow}

The authors are grateful to M.I.Dyakonov for very useful
discussions. This work was supported in part by
the Swedish Royal Academy of Science (grant 1240)
 and by the
Russian Foundation for Basic Research (Grant 96-02-17896).  Partial
support for one of the authors (V.Yu.K.) was provided
by a fellowship of INTAS Grant 93-2492-e within the research
program of International Center for Fundamental Physics in
Moscow. One of the authors (I.V.G.) is grateful to
ISSEP for Soros Postgraduate Student Grant.
This work was also supported by Grant 1001 within
the program "Physics of Solid State Nanostructures".
Two of us (A.P.D. and V.Yu.K) express gratitude to Uppsala
University for hospitality.

\section { Appendix A}
\label{App:A}

The conductivity correction corresponding to the diagrams (1a)
is given by
\begin{equation}
\Delta\sigma_{a} = -\sum^{\infty}_{N=3}
\frac {e^2 \hbar^3 }{2\pi m^2 S}
 \gamma
\int d^2 \br_i d^2 \br_f   \frac{\partial}{\partial \br_i}
 G^R_{i1}\frac{\partial}{\partial \br_f} G^A_{f1}
\Gamma_{N-1}
G^R_{fN}G^A_{iN}
d^2 \br_1...d^2 \br_N ,
\label{A1}
\end{equation}
where
$$
\Gamma_{N-1}=\left(  \gamma^{N-1}
G^R_{12}G^A_{12}...G^R_{N-1,N}G^A_{N-1,N} \right).
$$
Using Eqs. (\ref{f6}) and (\ref{f8}) we rewrite
the expression (\ref{A1}) as

\begin{equation}
\Delta\sigma_{a} = -\sum^{\infty}_{N=3}
\frac {e^2 \hbar^3 }{2\pi m^2 S}
 \gamma
\int d^2 \br_i d^2 \br_f G^A_{iN} \frac{\partial}{\partial \br_i}G^R_{i1}
G^R_{fN} \frac{\partial}{\partial \br_f} G^A_{f1}
W_{N-1}(\br_1-\br_N)
d^2 \br_1 d^2 \br_N.
\label{A2}
\end{equation}

Using Eq. (\ref{f9}) for
integration over $\br_i, \br_f$ in Eq. (\ref{A2})
we obtain
$$
\Delta\sigma_{a}
=-\sum^{\infty}_{N=3} \frac{e^2 n \tau}{mS\pi} (\lambda_F l)
\int d^2\br_1d^2\br_2d^2\br_N
P(\br_1-\br_2)W_{N-2}(\br_2-\br_N)P(\br_N-\br_1).
$$
Here we neglect the rapidly oscillating products
$G^RG^R$ and $G^AG^A$.
Finally, using Eqs. (\ref{f8},\ref{f10}) we derive
Eq. (\ref{coop}) presented in the main text.

\section { Appendix B}
\label{App:B}

The conductivity correction corresponding to the diagrams (1b)
is given by
\begin{equation}
\Delta\sigma_{b} = -2\sum^{\infty}_{N=3}
\frac {e^2 \hbar^3 }{2\pi m^2 S}
\gamma^2
\int d^2 \br_i d^2 \br_f   \frac{\partial}{\partial \br_i}
 G^R_{i1}\frac{\partial}{\partial \br_f} G^A_{f2}G^R_{12}
\Gamma_{N-2}
G^R_{N1}G^A_{iN}G^R_{1f}
d^2 \br_1...d^2 \br_N,
\label{B3}
\end{equation}
$$
\Gamma_{N-2}=\left(  \gamma^{N-2}
G^R_{23}G^A_{23}...G^R_{N-1,N}G^A_{N-1,N}\right).
$$

The factor $2$ in the Eq. (\ref{B3})
arises due to the consideration of both diagrams (1b)
which are complex
conjugated to each other.
Using (\ref{f6}) and (\ref{f8}) we rewrite
the expression (\ref{B3}) as
\begin{equation}
\Delta\sigma_{b} = -2\sum^{\infty}_{N=3}
\frac {e^2 \hbar^3 }{2\pi m^2 S}
\gamma^2
\int d^2 \br_i d^2 \br_f G^A_{iN} 
\frac{\partial}{\partial \br_i}G^R_{i1}
G^R_{12}G^R_{N1} G^R_{1f}
\frac{\partial}{\partial \br_f} G^A_{f2}
W_{N-2}(\br_2-\br_N)
d^2 \br_1d^2 \br_2 d^2 \br_N .
\label{B4}
\end{equation}
Using Eq. (\ref{f9}) and neglecting
the rapidly oscillating functions we get Eq. (\ref{nasha})
of the main text.

\newpage
\centerline{\bf FIGURES}

Fig. 1  Diagrams relevant in the first order in $(k_F l)^{-1}$:
the diagram describing coherent backscattering (a) and
the diagrams describing coherent scattering at
different angles. The contribution of the diagrams
(b) depends on the magnetic field. The contributions
of diagrams of the types (c) and (d) compensate each other.

Fig. 2  The angle dependence of the modified differential
cross-section on single impurity

Fig. 3  The process related to the diagrams (1a).
(a)  key points (i, f, 1, and N) have arbitrary positions;
(b) the positions  of key points satisfy the stationary
phase condition

Fig. 4  Similar to Fig. 4, but for the case of the diagrams (1b)

Fig. 5 The conductivity correction dependence on the
magnetic field at $\tau_{\phi}=\infty$.
The contributions of the diagrams (1a) also is presented.

Fig. 6  The conductivity correction dependence on the magnetic field
at different breaking phase times


\begin{thebibliography}{}
\bibitem{snoska} There are  corrections to the elastic scattering
time of the order of $(k_Fl)^{-1}$ and higher due to the renormalization
of the imaginary part of the self-energy operator. It is easy to calculate 
the correction in the first order in $(k_Fl)^{-1}$.
This correction, however, 
is not related to the weak localization effects, so that we don't
consider it in this paper.
\bibitem{two} The diagram (1a) with two dashed lines should not be
taken into account because in the first order in
$(k_Fl)^{-1}$ it is exactly compensated by the sum of two
complex conjugated diagrams (1b) with two dashed lines.
\bibitem{pi} In order to obtain the scattering
reduction  one should
add to $\Delta\phi$ an additional scattering phase acquired during the
first way after two collisions with impurity $1$ and
take into account the phase factor arising due to
integration over the impurity positions near the stationary
phase configuration. The resulting phase difference between
the two ways is equal to $\pi$.
\bibitem{valya} 
In frame of our interpretation it would be more
consistent to define the conductivity corrections due to 
diagrams (1a) and (1b) as
$$
\delta\sigma_{a,b}=-\sigma_0(2\pi S_0)^{-1}
\int^{2\pi}_0\Delta S_{a,b}(\theta)(1-\cos{\theta})d\theta,
$$ 
which differ from $\Delta\sigma_{a,b}$ respectively.
However, it follows from Eq. (\ref{nol}) that
$\delta\sigma_a+\delta\sigma_b=
\Delta\sigma_a+\Delta\sigma_b$ and such a definition
leads to the same result (Eq. (\ref{total})) for the
total conductivity correction.
\end{thebibliography}
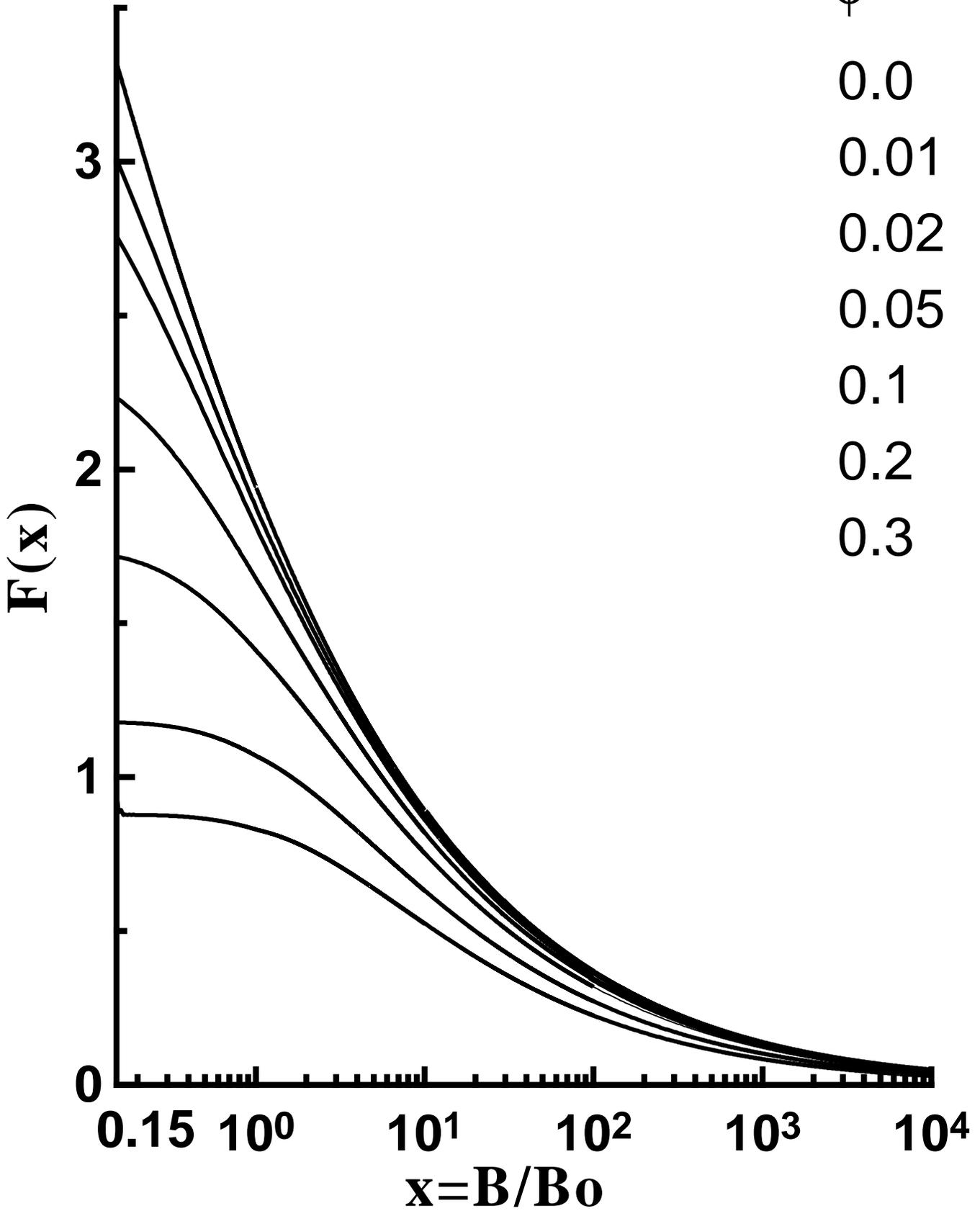
\end{document}